\documentclass{PoS}

\title{3D and 4D noncommutative electromagnetic duality and the role of the slowly varying fields limit}

\ShortTitle{3D and 4D NC EM duality and the role of the SVF limit}

\author{\speaker{Davi C. Rodrigues} and Clóvis Wotzasek\\
        Instituto de F\'{\i}sica, Universidade Federal do Rio de Janeiro, 21945-970, Rio de Janeiro, RJ, Brazil.\\
        E-mail: \email{cabral, clovis@if.ufrj.br}}

\abstract{We study classical noncommutative (NC) electromagnetic duality in both 3D and 4D  space-times through the Seiberg-Witten (SW) map to all orders in $\theta$. We evaluate the role of space-time dimensions, of the gauge coupling constant $g^2$ inversion, of the slowly varying fields (SVF) limit and of the rule $\theta \stackrel{\mbox{ \tiny duality}} \longrightarrow \tilde \theta = g^2 \; \str \theta$ (where $\str$ is the Hodge duality operator), which was originally found in the 4D space-time. Among our results, a new scalar picture for NC electromagnetism to second order in $\theta$ is established, a formula which simplifies considerably the application of the SW map in 3D is presented and we show that the SVF limit has a crucial role in this duality starting from the third order in $\theta$ for any dimension: outside this limit the symmetry between $\theta$ and $\tilde \theta$ is lost.}

\FullConference{Fifth International Conference on Mathematical Methods in Physics --- IC2006\\
		 April 24-28 2006\\
		 Centro Brasilerio de Pesquisas Fisicas, Rio de Janeiro, Brazil}

\def\[{\left\lbrack}
\def\]{\right\rbrack}

\def\({\left(}
\def\){\right)}
\newcommand{\be}{\begin{equation}}
\newcommand{\ee}{\end{equation}}
\newcommand{\ea}{\end{eqnarray}}
\newcommand{\ba}{\begin{eqnarray}}
\newcommand{\la}{\langle}
\newcommand{\ra}{\rangle}
\newcommand{\asp}{\textquotedblleft}  

\newcommand{\ep}{{\epsilon}}

\newcommand{\fdu}{{^{\star}F}}

\newcommand{\jdu}{{^{\star}J}}

\newcommand{\fmnd}{F_{\mu \nu}}

\newcommand{\prt}{{\partial}}
\newcommand{\diag}{\mbox{diag}}

\newcommand{\tr}{\mbox{tr}}
\newcommand{\tht}{\tilde \theta}
\newcommand{\hepth} {hep-th/}
\newcommand{\str}{{^\star}}
\newcommand{\real}{I \! \! R}

\begin{document}

\section{Introduction}

	
	The classical 4D electromagnetic duality without sources states an exchange between electric and magnetic fields $[ (\vec E, \vec B) \rightarrow (\vec B, -\vec E)$]. In regard to the electromagnetic waves this simply implies a $\pi / 2$ rotation, which, without sources, is physically innocuous. The situation is much more involving in nonlinear extensions of electromagnetism. Even without sources the NC electromagnetism is not a free theory and the extension of the mentioned duality leads to nontrivial properties essential to the understanding of the NC gauge theories. In this paper, which specially relies on Ref. \cite{issues}, we study some of the properties and nontrivial consequences of this duality within the NC field theoretical framework \footnote{From the string theory perspective, $S$-duality of $IIB$ strings in the presence of a magnetic background induces a duality between spacially NC Yang-Mills $\cal N$$=4$ theory  with a string model called NCOS (noncommutative open string), as conjectured in Ref. \cite{stringd}. Although the approaches of Ref. \cite{stringd} and ours are quite different, there are similarities in the resulting dualities, like the exchange of $\theta$ with $ {^\star} \theta g^2$. See our Conclusions for further comments.}.
	
	
	The first paper to address the issue of 4D NC electromagnetic duality \cite{ganor} employed the Seiberg-Witten (SW) map \cite{sw} to first order in $\theta$ and found that the dual action remains the same, except for the inversion of the coupling constant and the exchange of $\theta$ (the noncommutativity parameter) with $\tilde \theta = g^2 \;\str \theta$ (where $g^2$ is the gauge coupling constant and $\str$ is the Hodge duality operator). This modification on $\theta$ connects space-like noncommutativity with time-like noncommutativity $(\theta^{\mu \nu} \theta_{\mu \nu} > 0 \Rightarrow  \tilde \theta^{\mu \nu} \tilde \theta_{\mu \nu}< 0)$ and hence, if $\theta^{0 i}=0$, in the dual picture a noncommutativity between time and space emerges, a commonly undesirable feature\cite{uni}. Qualitatively, having in mind the $\pi/2$ rotation of electromagnetic waves in the ordinary case and that $\theta$ in general determines two preferential directions in space ($\theta^{0i}$ and $\ep^{ijk} \theta_{jk}$), which interacts with $\vec E$ and $\vec B$ differently, it should be clear that in the NC duality case the rule $ F \rightarrow \str F$ alone should not lead to another physically equivalent picture in general, some modification in $\theta$ is indeed necessary to assure duality. 

The result of Ref. \cite{ganor} was extended by Ref. \cite{aschieri} to all orders in $\theta$ in the slowly varying fields (SVF) limit \cite{sw, esw}, that is, disregarding possible appearence of derivative corrections on $F$ in the effective SW mapped action ($S = S|_{\mbox{\tiny SVF}} + O (\prt F)$). It was found that the rule $\theta \rightarrow \tilde \theta$ persists in that limit to all orders in $\theta$.

Many papers have addressed the NC extension of the 3D electromagnetic duality with topological mass \cite{topmas, dayi,  nosso, hr} which is a vector/vector duality. The 3D duality we study in this paper, likewise in Ref. \cite{issues}, is qualitatively different since it is a vector/scalar duality, which to our knowledge was not explored in a NC context before. Since it does not have topological mass it is easier to compare with the 4D duality.

In this work we extend the 4D NC electromagnetic duality to the 3D space-time and evaluate the necessity of the SVF limit from a classical field theoretical perspective, in order to find what are the  fundamental properties of this duality. Many arguments of Ref. \cite{ganor} depend on the space-time dimension (e.g., the 4D space-time is the only one which $\theta$ and $^\star \theta$ are both 2-forms and the $S$-dual massless gauge fields are both 1-forms), therefore a natural question is how the NC electromagnetic duality presents itself in other dimensions, and to what extent the properties of the 4D NC electromagnetic duality can be extended to those. From all possibilities, the 3D space-time seems to be a natural option. In this space-time, we establish to second order in $\theta$ the dual scalar action (consistently with the rule $\theta \rightarrow \, ^\star \theta g^2$) and we show that many terms of the Seiberg-Witten mapped action can be considerably simplified. The necessity of the SVF limit, to preserve the rule $\theta \rightarrow \, ^\star \theta g^2$ and therefore $S$-duality\footnote{In the sense of a global inversion of the coupling constant (string $S$-duality is not of concern in this approach).}, starts from the third order in $\theta$ for any space-time dimension (with $D \ge 2$).



\vspace{.5in}
\section{Revisiting the 3D electromagnetic duality}

To introduce our framework, we briefly review the electromagnetic duality in 3D ordinary space-time. The electromagnetic theory action with a 1-form source $J$ is
\be
	\label{e1}
	S_A[A,J] = \int \( a \; F \wedge \fdu \; + \; e \; A \wedge \jdu \),
\ee
where $A$ is the 1-form potential, the field strength $F$ satisfies, by definition, $F = dA$ and $a = -1/(2g^2)$. To preserve gauge invariance and to satisfy the continuity equation, $\jdu$ must be a closed 2-form.

As usual, the dynamics of the electromagnetic fields comes from the equation of motion and the Bianchi identity, namely,
\be
	\label{em1}
	d \fdu = - \frac e {2a} \; \jdu \;\;\;\;\;\;\; \mbox{and} \;\;\;\;\;\;\; dF = 0.
\ee 
Except for the sign of the first equality, the above equations are valid in any space-time dimension. One may introduce an electric vector $\vec E$ and a magnetic pseudo-scalar $B$ in 2D space and search for duality in these grounds \cite{issues}, but here we will follow the master action (or Lagrangian) approach\cite{master}. Consider the action
\be
	\label{m1}
	S_{M}[F, \phi] = \int \[ a F \wedge \( \fdu + \frac {e} a \Lambda \) -  d \phi \wedge F \],
\ee
where $F$, $\Lambda$ and $\phi$ are regarded as independent 2-form, 1-form and 0-form respectively, and $\Lambda$ is not a dynamical field. Equating to zero the variation of the above action with respect to $\phi$,  we obtain $dF =0$. This implies, in Minkowski space, that $F=dA$. Replacing $F$ by $dA$ and setting $\jdu = d \Lambda$, $S_M$ becomes equivalent to the action in Eq. (\ref{e1}). 

On the other hand, the variation of Eq.(\ref{m1}) with respect to $F$ produces
\be
	\label{map2}
	\fdu =  \frac 1 {2a} \( d\phi - e \Lambda \).
\ee 

Inserting Eq.(\ref{map2}) into the master action $S_M$ (and recalling ${^\star}{^\star} =1$ for any differential form in the dealt space-time), we find

\be
	\label{sp}
	S_{M} [F, \phi] \leftrightarrow - \frac 1 {4a} \int (d\phi - e\Lambda) \wedge {^\star}(d\phi - e\Lambda) 	= S_{\phi_\Lambda} [\phi].  
\ee
	
We use the symbol \asp $\leftrightarrow$" instead of \asp $=$" to be clear that equivalence of actions (functionals) is to be understood as a correspondence between their equations of motion; that is, if $ S_1 \leftrightarrow S_2$, the set of equations of $S_1$ can be manipulated, using its own equalities, or inserting new redundant ones, to become the set of equations of $S_2$ (the inverse also proceeds). 

The two equations of motion of $S_M$ ($dF = 0$ and Eq.(\ref{map2})) generate a map between the equations of motion of $S_A$ and $S_\phi$, viz,
\be
	{^\star dA} =  \frac 1 {2a} \( d\phi - e \Lambda \).
\ee 
Applying $d$ on both sides, we find Eq.(\ref{em1}), while the application of $d^\star$ results in  $d^\star d \phi =  e \, d^\star \Lambda$, which is the equation of motion of $S_\phi$.

\vspace{.5in}

\section{3D NC  electromagnetic duality to second order in $\theta$}

The NC version of the $U(1)$ gauge theory, whose gauge group we denote by $U_*(1)$, is given by \cite{rev}
\be
	\label{sm}
	S_{\hat A*} = a \int \hat F \wedge_*  {^\star} \hat F ,
\ee
where $a$ is a constant, $\hat F = d \hat A - i \hat A \wedge_* \hat A = \frac 12 (\prt_\mu \hat A_\nu - \prt_\nu \hat A_\mu - i[\hat A_\mu, \hat A_\nu]_* ) d x^\mu \wedge dx^\nu$, $[A,B]_* = A*B - B*A$ and
\be
	(A * B)(x) = \exp \( \frac i 2 \theta^{\mu \nu} \prt^x_\mu \prt^y_\nu \) A(x) \; B(y)|_{y \rightarrow x}
\ee
is the Moyal product. In particular, $[x^\mu, x^\nu]_* = i \theta^{\mu \nu}$. $(\theta^{\mu \nu})$ can be any real and constant anti-symmetric matrix.

Since $d \hat F \not= 0$, previous duality arguments cannot be directly applied. In order to employ them we will resort to the Seiberg-Witten (SW) map. Seiberg and Witten have shown that there must exist a map between $U_*(N)$ and $U(N)$ gauge theories \cite{sw}; therefore, denoting their gauge fields by $\hat A$ and $A$ respectively, it should comply with
\be
 \delta_{\hat \lambda} \hat A (A) = \hat A(A + \delta_\lambda A) - \hat A(A),
\ee
where $\hat A \in u_*(N)$ and $A \in u(N)$, i.e., for $N=1$, $\hat A$ transforms as $\delta_{\hat \lambda} \hat A = d \hat \lambda - 2 i \hat A \wedge_* \hat \lambda$ and  $A$ as $\delta_\lambda A = d \lambda$. While the $U_*(1)$ theory depends on $\theta$ only through the Moyal products, the mapped $U(1)$ theory only uses ordinary products and depends explicitly on $\theta$. As a corollary, also useful to our purposes, this map provides a more direct treatment of the observables \cite{jackiw}. While $\hat F \longrightarrow S_* * \hat F * S_*^\dagger$ under gauge transformations ($S_* \in U_*(1)$), $F$ is gauge invariant: $ F \longrightarrow S \;  F \; S^\dagger = F$ ($S \in U(1)$). In short, the mapped $U(1)$ theory satisfies the identity $d F =0$ and $F$ is a observable.

To second order in $\theta$, for the $U(1)$ case, the SW mapped action of (\ref{sm}) reads\footnote{Note that Ref.\cite{ganor} uses a different convention in the differential forms constant factors.} \cite{second, ganor, dayi}
\be
	\label{a*2}
	S_{A_\theta} = \frac a2 \int \[F^{\mu \nu} F_{\mu \nu} \( 1 - \frac 12 \theta^{\mu \nu} F_{\mu \nu} \) + 2 \; \tr (FF\theta F)+ L_{\theta^2} \] d^Dx,
\ee
with $D$ the space-time dimension,
\ba
	L_{\theta^2} =&& -2 \; \tr (\theta F \theta F^3 ) + \tr (\theta F^2 \theta F^2) + \tr (\theta F) \; \tr(\theta F^3) - \frac 18 \tr (\theta F)^2 \; \tr (F^2) + \nonumber \\
	&& + \frac 14 \tr(\theta F \theta F) \; \tr (F^2)
\ea
and $\tr (AB) = A_{\mu \nu} B^{\nu \mu}$, $\tr (ABCD) = A_{\mu \nu} B^{\nu \lambda} C_{\lambda \kappa} D^{\kappa \mu} \;$ \emph{etc}.

Fortunately, in the 3D space-time, the above expression can be considerably simplified. A direct computation shows that  $\tr (FF \theta F) = \frac 12  \tr (FF) \; \tr (F \theta)$ for any $F$ and $\theta$. With some reflection this relation can be generalized to 
\be
	\label{rel}
	\tr (A B_1 A B_2 \; ... \; A B_n) = \( \frac 12 \)^{n-1} \prod^n_{k=1} \tr (A B_k),
\ee
for any anti-symmetric 2-rank tensors $A, \{ B_k \}$. Therefore,
\be
	\label{ls}
	L_{\theta^2} = \frac 14 \tr (FF) \; \tr( \theta F)^2.
\ee




Exploiting the Bianchi identity, we propose the following master action to second order in $\theta$:
\be
	\label{amt}
	S_{M_\theta}[F, \phi] = \int \[ a \, \fdu \wedge F \, \( 1 + \langle \theta, F \rangle - \langle \theta, F \rangle^2 \)  - d \phi \wedge F \],
\ee
where $ \la \; , \, \ra$ is the scalar product between differential forms\footnote{In odd dimensional Minkowski space, the internal product of two n-forms $A$ and $B$ is defined by $\la A, B \ra = {^\star ( {^\star A} \wedge B)} = \frac 1{n!} A_{\mu_1 \; ... \; \mu_n} B^{\mu_1 \; ... \; \mu_n}$.}, in particular $\la F, \theta \ra = {^\star}( \fdu \wedge \theta) = \frac 12 \theta^{\mu \nu} F_{\mu \nu}$.	Other master actions are possible \cite{issues}.

	The variation of $S_{M_\theta}[F, \phi]$ in  respect to $\phi$ leads to $dF=0$, which implies $F=dA$; inserting this result into $S_{M_\theta}$, $S_{A_\theta}$ is obtained.	To settle the other side of duality, the variation in regard to $F$ is evaluated, leading to a nontrivial NC extension of Eq.(\ref{map2}) without source, namely,
\ba
	^\star F = && \frac {d \phi} {2a} \( 1 - \frac {\la \theta, {^\star d \phi} \ra}{2a} - 3 \frac {\la \theta, { ^\star d \phi} \ra^2}{ 4a^2} + \la \theta, \theta \ra \frac {\la d \phi, d \phi \ra }{ 8a^2} \) - \nonumber \\
	\label{fm}
	&& - {^\star} \theta \frac {\la d \phi,  d \phi \ra }{8 a^2} \( 1 -  5 \frac {\la \theta, {^\star} d \phi \ra} {2a} \).
\ea

The scalar action which describes NC electromagnetism to second order in $\theta$ then reads
\be
	\label{p*2}
	S_{\phi_\theta} = - \frac 1 {4a} \int d \phi \wedge {^\star} d \phi \( 1 - \la  \tht, d \phi \ra + 3 \la  \tht, d \phi \ra^2 + \frac 14 \la \tht, \tht \ra \; \la d \phi, d \phi \ra \),
\ee
where $\tht = {^\star} \theta / 2a$.

The correspondence of the equations of motion between vector and scalar models, as expected, is given by $F=dA$ together with Eq.(\ref{fm}) (and its inverse). 

This duality to second order in $\theta$ shows the exchange of the coupling constant $a$ with its inverse and  that this inversion only occurs if a certain rescaling of $\theta$ is assumed, which happens if $\theta \longrightarrow \tilde \theta$. This rule is the one found when studying the S-duality of NC field theories in 4D space-time in the following contexts: to first order in $\theta$ \cite{ganor}, in the slowly varying fields limit \cite{aschieri} and as a consequence of the IIB string S-duality \cite{stringd}. In the 3D space-time in particular it rotates the  privileged spacial direction by\footnote{In the 3D space-time $\theta$ determines a single privileged direction in space given by $(\vec \theta)^i = \theta^{0i}$, which is orthogonal to the privileged direction given by $\str \theta$.} $\pi /2$. If no spacial privileged direction is present ($\theta^{0i}=0$), then duality preserves spacial isotropy.

The action (\ref{p*2}) presents the 3D NC electromagnetism in the scalar picture up to second order in $\theta$ and shows that the rule $\theta \rightarrow \tilde \theta$ plays an important role in 3D NC electromagnetism duality in regard to classical S-duality. In the next section we investigate to what extent the interchange between $\theta$ and $\tilde \theta$ through duality is exact.



%

\vspace{.5in}

\section{Third and higher order duality and the role of the SVF limit}

NC gauge theories have a very interesting connection with general relativity, for coordinate transformations are achieved through gauge transformations (see \cite{szabonew} for a recent review). Nevertheless the SW map states an equivalence between NC gauge theories and ordinary gauge theories. The correspondence of the last with general relativity can be recovered by assuming that the $\theta$ dependent terms induced by the SW map are nonlinear-field-dependent metric corrections \cite{gravriv}. The realization of this idea to any order in $\theta$, in the SVF limit, was done in Ref. \cite{esw}. According with this approach, the SW map establish an equivalence between the NC electromagnetic action $\int \hat F \wedge_* \str \hat F$ with a certain commutative electromagnetism in a curved space-time which reads $ \int F \wedge  \hat \str F$, where $\hat \str$ stands for a Hodge duality operator evaluated with a metric which depends on $F$ and\footnote{In component notation this reads \cite{esw} $	\int \sqrt{| g|} \; \; g^{\mu \nu} \;   g^{\alpha \beta} \; F_{\mu \alpha} \; F_{\nu \beta}$ with $g_{\mu \nu} = \eta_{\mu \nu} + (F \theta)_{\mu \nu}$, $g = \det(g_{\mu \nu})$ and $\eta = \diag(+ - - ...)$.} $\theta$. 

In the SVF limit, to any order in $\theta$ and in any space-time dimension, no contribution of $\prt F$ appears in the mapped action. Moreover, since the SW mapped action must be invariant under $U(1)$ gauge transformations\footnote{And its structure is not dimensional dependent.} the gauge potentials $A$'s should only appear inside the field strengths $F$'s. Therefore, applying formula (\ref{rel}), in the 3D space-time the general term of the SW map can be written as 
\be
	\label{general}
	a \; c_n \int F \wedge \str F \; {\la F, \theta \ra}^n,
\ee 
where $c_n \in \real$.

When evaluating 3D duality, one should notice that, since the term of n${^{\mbox{\small th}}}$ order in $F$ occurs in (\ref{general}) proportionally to $\theta^{n-2}$, for each $\phi$ which appears in the dual scalar action there will correspond a factor $1/a$ and in the term of n$^{\mbox{\small th}}$ power in $\phi$ there will be a term of $(n-2)^{\mbox{\small th}}$ order in $\theta$; therefore the rule $\theta \longrightarrow \tilde \theta = \str \theta /2 a$ will assure a global inversion of the coupling constant, likewise previously stated up to the second order in $\theta$. Furthermore it should be clear that, to any order in $\theta$, this rule is such that in the dual picture no explicit reference to $\theta$, $\str \theta$ or $\str \tilde \theta$ is necessary, $\tilde \theta$ can be seen as the fundamental parameter of the dual picture. Therefore in the 3D space-time with the SVF limit the rule of interchanging $\theta$ with $\tilde \theta$ through duality remains exact to any order in $\theta$, likewise in the 4D case. Additionally, one may write the dual scalar action as $\int d \phi \wedge \tilde \str d \phi$, where $\tilde \str$ is an extension of the Hodge duality operator in 3D space-time which depends on $d \phi$ and $\tilde \theta$.

\vspace{.2in}


A natural question that emerges here is if the SVF limit has any role in the study of this duality. Up to the second order in $\theta$ no contribution of $\prt F$ appears in the SW mapped action. However, as we will show, starting from the third order expansion in $\theta$, terms with more derivatives than potentials appear in the SW map of $\hat F$ and are present in $L_{\theta^3}$. These factors spoil the last suggested symmetry between $\theta$ and $\tilde \theta$. To infer these terms, we will use the following SW differential equation \cite{sw}
\ba
	 \label{dsw}
	 \delta \hat \fmnd	(\theta) &=& \frac 14 \delta \theta^{\alpha \beta} \[ 2 \hat F_{\mu \alpha} * \hat F_{\nu \beta} + 	2 \hat  F_{\nu \beta} * \hat F_{\mu \alpha}  - \hat A_\alpha * (\hat D_\beta \hat F_{\mu \nu} + \partial_\beta \hat F_{\mu \nu})  - \right. \\ \nonumber
	&&  \left. - (\hat D_\beta \hat F_{\mu \nu} + 	\partial_\beta \hat F_{\mu \nu}) * \hat A_\alpha \].
\ea

Expanding $\hat F$ and $\hat A$ in powers of $\theta$, to third order it reads
\be
	 \label{dsw3}
	 \delta \hat \fmnd^{(3)}	(\theta) = - \frac 14 \delta \theta^{\alpha \beta} \theta^{\alpha' \beta'}\theta^{\alpha'' \beta''} \( \prt_{\alpha'}\prt_{\alpha''}F_{\mu \alpha} \prt_{\beta'}\prt_{\beta''}F_{\nu \beta} - \prt_{\alpha'}\prt_{\alpha''} A _\alpha \prt_{\beta'}\prt_{\beta''} \prt_\beta F_{\mu \nu} \) +...
\ee
Where $F_{\mu \nu} = \hat F_{\mu \nu}^{(0)}$ and $A_\mu = A^{(0)}_\mu$. Only the terms with more derivatives than fields were written in the above expression. Inserting this result into Eq.(\ref{sm}), the only terms of $L_{\theta^3}$ which have more derivatives than fields are in the following expression\footnote{This solution can also be inferred by the results of Ref.\cite{fidanza}, Section 3.2, in which the SW map is expanded in powers of $A$.}
\be
	\label{df}
	\theta^{\alpha \beta} \theta^{\alpha' \beta'} \tr(\prt_{\alpha} \prt_{\alpha'} F \; \theta \; \prt_{\beta} \prt_{\beta'} F \; F) - \frac 14 \theta^{\alpha \beta} \theta^{\alpha' \beta'} \; \tr(F \theta) \; \tr (\prt_{\alpha} \prt_{\alpha'} F \; \prt_{\beta} \prt_{\beta'} F). 
\ee

Once again, in the 3D space-time a considerable simplification is possible. In the 3D space-time, the expression (\ref{df}) is equal to
\be
	\label{df3d}
	\frac 14 \; \theta^{\alpha \beta} \theta^{\alpha' \beta'} \; \tr( \prt_\alpha \prt_{\alpha'} F \; \prt_\beta \prt_{\beta'} F) \; \tr(F \theta).
\ee 

A careful analysis shows that (\ref{df}) [likewise (\ref{df3d})] is not identically null nor is a surface term in any space-time with dimension greater than one \cite{issues}. This result is in contradiction with a certain proposition of Ref. \cite{bc}, see our Conclusions for more details.

Since $F = \str d\phi / (2a) + O(\theta)$, to third order in $\theta$ the contribution of (\ref{df3d}) to the scalar action reads 
\be
	\int d^3x \; \frac a{2(2a)^3} \; \theta^{\alpha \beta} \theta^{\alpha' \beta'} \theta^{\nu \lambda} \; (\prt_\alpha \prt_{\alpha'} \prt_\mu \phi) \; (\prt_\beta \prt_{\beta'} \prt^\mu \phi) \; \ep_{\nu \lambda \rho} \prt^\rho \phi.
\ee
This term violates the symmetry between $\theta$ and $\tilde \theta$. It is natural to change the third $\theta$ above to $\tilde \theta$, which will lead to preservation of the classical $S$-duality, nevertheless this transformation is rather artificial and leaves the dual picture as dependent of $\theta$ and $\tilde \theta$, which, in particular, for $\theta^{0i} \not=0$ will insert a new privileged direction in the dual picture. On the other hand, changing all the above three $\theta$'s to $\tilde \theta$'s will spoil the global inversion of the coupling constant. It is easy to see that similar arguments are valid to the 4D space-time.

\vspace{.5in}

\section{Conclusions}

We established, to second order in $\theta$,  the scalar description of the 3D NC electromagnetic theory (\ref{p*2}), which is usually described by the gauge model (\ref{sm}). In order to achieve this result, and some subsequent ones, we found and employed the formula (\ref{rel}), which significantly simplifies the Seiberg-Witten (SW) mapped action in 3D space-time. The rule $\theta \stackrel{\mbox{ \tiny duality}} \longrightarrow \tilde \theta = g^2 \; \str \theta$ was extended to the 3D space-time up to the second  order in $\theta$ or to all orders in $\theta$ in the slowly varying fields (SVF) limit; outside this limit and starting from the third order expansion in $\theta$, as shown in the last section, this rule is incompatible with classical $S$-duality.

Currents can be easily inserted in this duality, along the lines of Sec.2, if it is assumed a $\theta$ non-dependent coupling like $A \wedge {^\star J}$ in the mapped action. Nevertheless, this violates correspondence with the $U_*(1)$ theory, which asserts the coupling $\hat A \wedge_* {^\star} \hat J$, whose map was found in Ref.\cite{banerjee}.

In the previous section we show, by means of a straightforward calculation, that the SW differential equation (\ref{dsw}) leads to the appearance of terms with more derivatives than fields in the third order expansion. These terms were applied into the NC electromagnetic Lagrangian ($L_{\hat A_*})$ and the resulting terms were stated in (\ref{df}). Perhaps surprisingly, these terms are not null nor are surface terms, as carefully shown in Ref. \cite{issues}\footnote{In general the Seiberg-Witten map is not unique\cite{asakawa}, nevertheless the additional terms do not influence this analysis.}.  This result is not in agreement with the first part of a proposition in Ref.\cite{bc}. We think our result should be considered as a counter-example to it. Indeed, the first part of Proposition 3.1 does not seem to be correct in general \cite{comment}. However, it should be stressed that it clearly holds in the slowly varying fields limit and, in this limit, it is compatible with our results; moreover, any results which depend on that proposition are perfectly valid in that limit. 

The SW map can be used as a convenient tool to extract local gauge invariant quantities from the  NC gauge field theories \cite{jackiw}. An interesting question is if in general the SW map removes all the nonlocality of the original NC gauge theory. Our answer is no. Outside the SVF limit according with an immediate extension of our previous Section, there appears in the action terms whose number of derivatives is proportional to the power of $\theta$, so only disregarding terms with $\prt F$ one can achieve a local action exact in $\theta$ obtained through the SW map. 

As previously stated, this work does not aim to resolve string $S$-duality issues in the presence of a magnetic background, like Ref. \cite{stringd} does. However, a certain exchange of $\theta$ with ${^\star} \theta g^2$, among other similarities, occurs in both cases. According with our result, this exchange only occurs to all orders in $\theta$ in the SVF limit. At the moment it is not clear to us if our result has consequences to the string  $S$-duality of NC theories since, among other possibilities, we may have come across a pathological feature of the Seiberg-Witten map \cite{dm}.

\acknowledgments

This work is partially supported by PRONEX/CNPq. DCR thanks FAPERJ (Brazilian research agency) for financial support. We thank Prof. R. Banerjee and Prof. V. Rivelles for useful comments and Prof. S.L. Cacciatori and collaborators for their attention to our problem and their cordial answer.


\end{document}